\documentclass[a4paper,nofootinbib,showpacs,preprintnumbers,floatfix,superscriptaddress,prl,twocolumn,aps]{revtex4}
\usepackage{graphicx}
\usepackage{dcolumn}
\usepackage{amsmath}
\usepackage{amsfonts}
\usepackage[sans]{dsfont}
\usepackage{amsbsy}
\usepackage{color}
\usepackage{units}
\usepackage{bbm}
\usepackage{bm}
\usepackage[citebordercolor={0 0 1}, linkbordercolor={0 1 0}]{hyperref}
\newcommand{\proj}[1]{\ensuremath{|#1\rangle \langle #1|}}
\newcommand{\beq}{\begin{equation}}
\newcommand{\eeq}{\end{equation}}
\newcommand{\bea}[1]{\begin{equation}\begin{array}{#1}}
\newcommand{\eea}{\end{array}\end{equation}}
\newcommand{\beqn}{\begin{eqnarray}}
\newcommand{\eeqn}{\end{eqnarray}}
\newcommand{\ket}[1]{\ensuremath{\left|{#1}\right\rangle}}
\newcommand{\bra}[1]{\ensuremath{\left\langle{#1}\right |}}

\newcommand{\diad}[2]{\ensuremath{\left|{#1}\rangle\langle{#2}\right |}}

\newcommand{\U}{\textsf{U}}
\newcommand{\V}{\textsf{V}}
\newcommand{\T}{\textsf{T}}
\newcommand{\W}{\textsf{W}}
\newcommand{\A}{\textsf{A}}
\newcommand{\B}{\textsf{B}}
\newcommand{\F}{\mathcal{F}}
\newcommand{\tr}{\mathrm{Tr}}
\newcommand{\Abar}{\mathds{A}}
\renewcommand{\rho}{\varrho}

\begin{document}


\title{Optimal teleportation with a noisy source}

\author{B. G. Taketani}
\affiliation{Instituto de F\'{\i}sica, Universidade Federal do Rio de Janeiro, Rio de Janeiro, Brasil}
\affiliation{Physikalisches Institut der Albert--Ludwigs--Universit\"at, Freiburg, Deutschland}
\author{F. de Melo}
\affiliation{Instituut voor Theoretische Fysica, Katholieke Universiteit Leuven, Leuven, Belgi\"e}
\affiliation{Physikalisches Institut der Albert--Ludwigs--Universit\"at, Freiburg, Deutschland}
\author{R. L. de Matos Filho}
\affiliation{Instituto de F\'{\i}sica, Universidade Federal do Rio de Janeiro, Rio de Janeiro, Brasil}


\begin{abstract}
We establish the optimal quantum teleportation protocol for the realistic scenario when both input state and quantum channel are afflicted by noise.  In taking these effects into account higher fidelities are achieved. The optimality of the proposed protocol prevails even when restricted to a reduced set of generically available operations.
\end{abstract}


\maketitle


\emph{Introduction.} Information Theory's main concern is optimal data transmission over noisy channels. In his 1948 seminal article ``A Mathematical Theory of Communication"~\cite{SHANNON:1948fk}, C.~E.~Shannon set the foundation stone of this theory. He showed that below a certain transmission rate threshold, which depends on the amount of noise on the channel, there exists a data codification that enables transmission with asymptotically negligible error. Since only classical information was considered, the input could be assumed perfectly prepared with all the disturbances lumped into the transmission process.

\begin{figure} [!t]
\begin{center}
\includegraphics[width=\columnwidth]{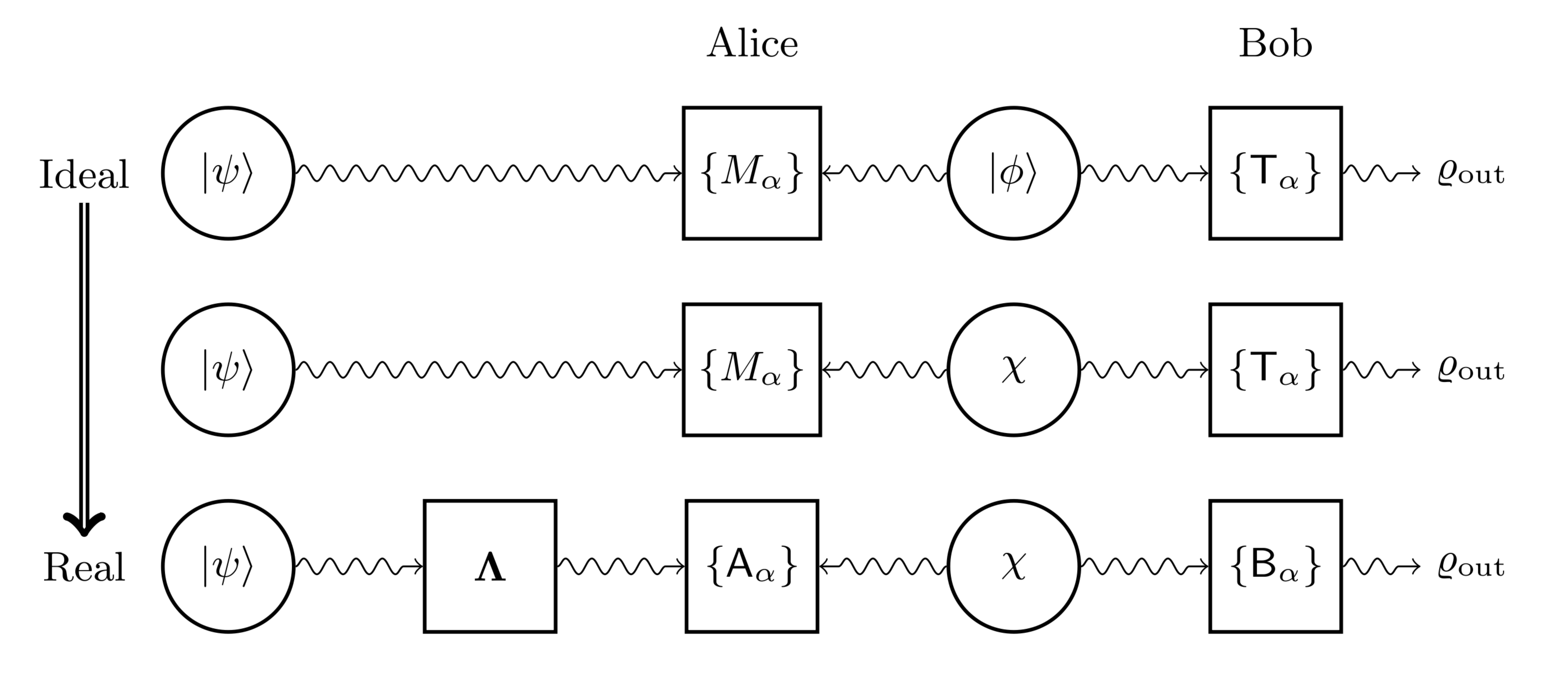}
\caption{{\bf  Teleportation protocol: from ideal to real.}  In the standard teleportation protocol (STP), first row, Alice and Bob  share a maximally entangled state $\ket{\phi}:=\sum_{i=0}^{n-1} \ket{ii}/\sqrt{n}$. The total initial state  $\ket{\psi}\otimes\ket{\phi}$ can be rewritten as $\sum_{\alpha} \ket{\phi_{\U_\alpha}}\otimes U_\alpha\ket{\psi}/n$, with $\ket{\phi_{\U_\alpha}}=\U_{\alpha}^{\dagger}\otimes\openone\ket{\phi}$ a ME state  and $\tr(U_\alpha^\dagger U_\beta)=n \delta_{\alpha,\beta}$.  Alice measures her two parties with projectors $M_\alpha= \proj{\phi_{\U_\alpha}}$. With probability $1/n^2$ she gets one of the ME states, and sends to Bob via a classical channel (not shown) its index $\alpha$. With this information, Bob performs a unitary transformation $T_\alpha = U_\alpha^\dagger$, and recovers the initial state without gaining any knowledge about it. In the second row, Alice and Bob share a non-maximally entangled state $\chi$. To maximize the protocol fidelity Alice and Bob optimize the measurement basis and unitary operations over the uniform distribution of initial pure states, for the source of input states is assumed noiseless. The realistic case in the third row, where noise is present in both the channel and source, and Alice and Bob are allowed to perform more general operations, is detailed in the text. 
\label{fig:diagrams}}
\end{center}
\end{figure}

The newly born theory of Quantum Information follows the same steps of its predecessor. A noisy channel theorem, in the same spirit as the one by Shannon, was proved for the quantum case by Holevo~\cite{Holevo:1998vn}, and Schumacher and Westmoreland~\cite{Schumacher:1997kx}. Information is now encoded into quantum bits (or qubits), and its transmission is via quantum channels. The quantum realm, however, presents a myriad of new possibilities, with the teleportation protocol being the most astonishing example. In the protocol devised by Bennett and co-authors~\cite{BENNETT:1993fk}, an unknown state is perfectly transmitted between two parties (usually dubbed Alice and Bob) with the aid of classical communication and a shared maximally entangled (ME) state -- the latter plays the role of a quantum channel, with no classical counterpart. See Fig.~\ref{fig:diagrams} for a brief review of the teleporation protocol. As in the classical case, idealized scenarios are quickly substituted by more realistic ones, and teleportation over noisy quantum channels has been an extensively investigated topic~\cite{Horodecki:1999cr,Albeverio:2002ys,Verstraete:2003zr,Bennett:1996uq,Modlawska:2008nx}. The action of the noise is represented by a (completely positive) map, that generically maps the shared initially pure maximally entangled state into a mixed state with less entanglement. The teleportation is no longer perfect. Since the input state is unknown, Alice and Bob optimize their actions such as to maximize the average protocol quality (fidelity) over the set of input states. One point, however, has been hitherto neglected: quantum information is unavoidably disturbed by the environment, even before its transmission through the channel. The proper averaging is thus not over the uniform distribution of pure input states, but over the initial input distribution induced by the environment. As observed in \cite{Henderson:2000ve}, {\it a priori} information about the distribution of states to be teleported can be used to achieve higher fidelities. Here we address this issue, and present the optimal teleportation protocol including the effect of a noisy source.  After that, we discuss the experimentally motivated scenario where Alice and Bob can implement only a small subset of all possible  physical operations. The gain of the proposed protocol in respect to  previous proposals is then numerically accessed. 

\emph{Realistic protocol.}
Let $\proj{\rm{\psi}}\otimes\chi$ be the total initial state.  $\proj{\rm{\psi}} \in \mathcal{H}_{in}$ is a unknown input state to be teleported, and $\chi\,\in\,\mathcal{H}_{A}\otimes\mathcal{H}_{B}$ is the noisy  quantum channel shared by Alice and Bob. For simplicity, we assume that  $\dim \mathcal{H}_A= \dim \mathcal{H}_{\rm{in}}=\dim \mathcal{H}_B=n$. However, as the initial state cannot be perfectly created, or is the product of previous processing, the actual state to be teleported is given as the result  of a completely positive map $\proj{\rm{\psi}} \mapsto \Lambda\left[\proj{\rm{\psi}}\right]$. The actual  state at hands is thus  $\Lambda \left[\proj{\rm{\psi}}\right]\otimes\chi$. 

 As in the standard teleportation protocol (STP), see Fig.~\ref{fig:diagrams}, Alice and Bob apply coordinated operations on their systems aiming for the highest teleportation fidelity. In general, this local, classically correlated  (LOCC) actions can be described by:
\begin{equation}
\Lambda \left[\proj{\rm{\psi}}\right]\otimes\chi \mapsto \sum_{\alpha}\big(\A_{\alpha}\otimes \B_{\alpha} \big) (\Lambda \left[\proj{\rm{\psi}}\right]\otimes\chi) \big(\A_{\alpha}^{\dagger}\otimes \B_{\alpha}^{\dagger} \big);\nonumber
\end{equation}
where $A_\alpha$ denotes Alice's operation on $\mathcal{H}_{\rm{in}}\otimes\mathcal{H}_A$, and  $B_\alpha$ represents Bob's reaction on $\mathcal{H}_B$~\cite{nota}. The common index $\alpha$ indicates the coordinated action, and is exchanged between Alice and Bob via a classical channel. In order to conserve probabilities, this operation must satisfy $\sum_{\alpha}\A_{\alpha}^{\dagger}\A_{\alpha}\otimes \B_{\alpha}^{\dagger}\B_{\alpha}=\openone\otimes\openone$. On avareage, Bob is left with the output state $\rho_{\rm{out}}$ given by
\begin{equation}
\sum_{\alpha}\tr_{\rm{in,A}}\left[\big(\A_{\alpha}\otimes \B_{\alpha} \big) (\Lambda \left[\proj{\rm{\psi}}\right]\otimes\chi) \big(\A_{\alpha}^{\dagger}\otimes \B_{\alpha}^{\dagger} \big)\right].
\label{eq:bob_out}
\end{equation}
This expression can be simplified noting that, by virtue of the Jamio\l kowsi~\cite{Jamiolkowski:1972qf} isomorphism, the noisy channel $\chi$ can be written as $\openone\otimes\bm\Gamma\left[\diad\phi\phi\right]$, with $\bm\Gamma[\bullet]:= \sum \Gamma_i \bullet \Gamma_i^\dagger$ a completely positive map, and $\ket{\phi}=\sum_{i=0}^{n-1} \ket{ii}/\sqrt{n}$ a maximally entangled state. It then follows that
\begin{align}
\rho_{\rm{out}}=\sum_{\alpha,i}\B_{\alpha}\,\Gamma_{i}
\Big\{\tr_{\rm{in},A}\Big[&\Big(\A_{\alpha}\otimes\openone\Big)(\bm\Lambda\left[\proj{\psi}\right]\otimes\proj{\phi})  \nonumber\\
&\times\Big(\A_{\alpha}^{\dagger}\otimes\openone\Big)\Big]\Big\}\Gamma_{i}^{\dagger}\,\B_{\alpha}^{\dagger}  .\nonumber
\label{sqt}
\end{align}
Expanding $\A_{\alpha}$ in a maximally entangled basis $\A_{\alpha}=\sum_{rs}a_{rs}^{\alpha}\diad{\phi_{\U_r}}{\phi_{\U_s}}$, where $\ket{\phi_{\U_\alpha}}=\U_{\alpha}^{\dagger}\otimes\openone\ket{\phi}$, and defining the operators $\mathds{A}^{r}_{\alpha}=\nicefrac{1}{n}\sum_{p}\bra{\phi_{\U_r}}\A_{\alpha}\ket{\phi_{\U_p}}\U_{p}$, we can write the output state as:
\begin{equation}
\rho_{\rm{out}}=\sum_{\alpha,r}\B_{\alpha}\,\bm\Gamma\big[\mathds{A}^{r}_{\alpha}
\;\bm\Lambda\left[\diad{\psi}{\psi}\right]\,\mathds{A}^{r \dagger}_{\alpha}\big]\,\B_{\alpha}^{\dagger}. \nonumber
\end{equation}
The most general teleportation protocol can thus be recast as the map $\bm\Phi: \mathcal{H}_{\rm{in}} \mapsto \mathcal{H}_B : \proj{\psi} \mapsto \rho_{\rm{out}}=\sum_k \Phi_k \proj{\psi} \Phi_k^\dagger$ with $\Phi_{k=\{\alpha,r,i,j\}} :=  \B_\alpha \Gamma_i \Abar_\alpha^{r} \Lambda_j$, where we used the decomposition for the map $\bm\Lambda[\bullet]=\sum_j \Lambda_j \bullet \Lambda_j^{\dagger}$. We can thus define an effective teleportation map as acting on the original pure state, and not on the mixed state which Alice actually manipulates. 

The still undefined operations $\{\Abar_\alpha^{r}\}$ and $\{\B_\alpha\}$ are to be fixed by optimizing the protocol for all possible input states. Even though the unknown system being teleported is in a mixed state, the primary goal of the protocol is to teleport the original pure state. Therefore, the figure of merit to be optimized is the average fidelity of the output state with the pure input state, i.e., $\overline{f}=\overline{\bra{\psi}\rho_{\rm{out}}\ket{\psi}}$. 

The evaluation of $\overline{f}$ is obtained by following the general framework develop in~\cite{Horodecki:1999cr}. The maximal average fidelity for the optimal protocol is then given by:
\begin{align}
\overline{f}_{\max}=\frac{n}{n+1}\F_{\max}\left(\chi,\bm\Lambda\right)+\frac{1}{n+1}\,\,\,,
\label{avefid}
\end{align}
where $\F_{\max}$, defined as
\begin{align}
\max_{\bm\Omega}\bra{\phi} (\openone\otimes\bm\Lambda)\circ\bm\Omega\left[\chi\right]\ket\phi ,
\label{Fmax}
\end{align}
is the maximal singlet fraction attainable by the combined action on $\chi$ of the trace-preserving operation $\bm\Omega$ and the decoherence map $\openone\otimes\bm\Lambda$. The maximization is taken over operations $\bm\Omega[\bullet] \mathrel{\mathop:}= \sum_{\alpha,r}\big(\Abar^{rT}_{\alpha}\otimes\B_{\alpha}\big)\bullet \big(\Abar^{rT}_{\alpha}\otimes\B_{\alpha}\big)^{\dagger}$, which refer to the LOCCs of Eq.~\eqref{eq:bob_out}. The transposition operation $T$ is taken on the computational basis. 
This concludes the protocol, which constitute our main result.


\emph{Discussion.}   To highlight the importance of acknowledging the presence of noise in the input distribution of states, we compare the protocol introduced above with the two main protocols for handling noisy teleportation.


\noindent \emph{i) Optimal teleportation vs. Distillation+STP.} In Ref.~\cite{Horodecki:1999cr}  the authors realized that, in the noiseless input case, the optimal teleportation protocol is equivalent (in the sense of average fidelity) to an optimal distillation of the resource state followed by a STP. Explicitly, when $\bm\Lambda=\openone$  we have that $\F_{\max}$ is
\beq
\max_{\bm\Omega}\bra{\phi} \bm\Omega\left[\chi\right]\ket\phi = \bra{\phi} \bm\Omega_\text{STP}\left[\rho_*\right]\ket\phi; \nonumber
\eeq
where  $\rho_*= \bm\Omega_* [\chi]$, with $\bm\Omega_*$ the optimal distillation, and $\bm\Omega_\text{STP} [\bullet]  = 1/n^2 \sum_\alpha \U_\alpha^T\otimes\U_\alpha^\dagger \bullet (\U_\alpha^T\otimes\U_\alpha^\dagger)^\dagger$ representing the standard teleportation protocol. This is easily realized by noting that the singlet fraction is invariant under the STP, i.e., $\bra{\phi} \bm\Omega_\text{STP}\left[\mathcal{O}\right]\ket\phi = \bra{\phi} \mathcal{O}\ket\phi $ for any $\mathcal{O}$. Therefore,  Eq.~\eqref{avefid} tell us that performing the STP with the optimal distilled state $\rho_*$ yields the same average fidelity as performing the optimal teleportation protocol $\bm\Omega_*$ with the original resource $\chi$. This equivalence was then used by Verstraete and Verschelde (VV)~\cite{Verstraete:2003zr}, to design an optimal teleportation protocol via the best distillation procedure.

This correspondence, however, breaks down for noisy input states. Although mathematically it still remains true that $\max_{\bm\Omega}\bra{\phi} (\openone\otimes \bm\Lambda)\circ\bm\Omega\left[\chi\right]\ket\phi = \bra{\phi} \bm\Omega_\text{STP}\left[\rho^\prime_*\right]\ket\phi$, where now $\rho^\prime_*= {\bm\Omega}^\prime_* [\chi]$ with ${\bm\Omega}^\prime_*$ the optimal operation in~\eqref{Fmax}, physically the equivalence would assume that the effect of the noise $\bm\Lambda$ in the input states can -- as in the classical paradigm --  be absorbed in the channel (resource state). For a sensible correspondence still to prevail in the noisy input scenario, we must require that $\bra{\phi} \bm\Omega_\text{STP} \circ (\openone\otimes\bm\Lambda)[{\rho}^\prime_*]\ket\phi = \bra{\phi} (\openone\otimes\bm\Lambda)\circ\bm\Omega_\text{STP}[{\rho}^\prime_*]\ket\phi$.  This is not true in general.

In this way, in a realistic scenario, the protocol proposed by VV is no longer the optimal one and must be replaced by the protocol here introduced. Operationally the reason for the latter  to be at least as good as the first is clear: the realistic protocol allows Alice to perform general, collective operations on both input and (half) resource state, which is obviously superior than acting only on the resource state as in the VV protocol, or even separately on input and resource states.


\noindent \emph{ii) Unitaries+projective~measurements.} 
The realistic protocol, \eqref{avefid} and \eqref{Fmax}, supposes the ability to perform the most general operations on Alice and Bob's parties. This may be impracticable. The most general LOCC operation may, for instance, require an infinite amount of classical communication exchange. The optimization of the protocol is thus only defined given an specific experimental realization and the accessible operations at the moment. A trade-off between protocol quality and experiment complexity should be always observed.  Arguably the simplest protocol is the one  where Alice performs projective measurements on a maximally entangled basis $\{\proj{\phi_{\U_{\alpha}}}\}$, and Bob applies unitary transformations $\{\T_{\alpha}\}$ depending on the measurements outcome. Within these operations we can exactly pinpoint the advantage of taking into account the noise on the source, for~\eqref{Fmax} reduces to
\beq
\F_{\max}=  \frac{1}{n^{2}}\!\max_{\{\U_\alpha,\T_\alpha\}}\sum_{\alpha , k , l}|\bra{\phi} \Lambda_k^T \otimes \T_\alpha\Gamma_l\U_\alpha\ket{\phi}|^2;
\label{febf}
\eeq
with the optimization taken over all unitary basis $\{\U_\alpha\}$, and all sets of unitary matrices $\{\T_\alpha\}$.
Here again, it is clear that the case where the noise in the source is not taken into account is far from general. In fact, by setting $\Lambda_k \propto \openone$ each term in~\eqref{febf} is equal to $\sum_{l}|\bra{\phi} \openone \otimes \V_\alpha \Gamma_l\ket{\phi}|^2$, with $V_\alpha = \U_\alpha\T_\alpha$. As the $\T_\alpha$'s are not subjected to any constraint, each of these terms can be optimized independently over $\V_{\alpha}$'s. The optimal fidelity is thus obtained for any choice of measurement basis. This simplified case was obtained in Ref.~\cite{Albeverio:2002ys}. When $\Lambda_k \not\propto \openone$, the optimization is much more challenging as each term in the sum is ``coupled" to the others via the unitary basis constraint.

Another interesting scenario is recovered when the map $\bf\Gamma$ (and/or $\bf \Lambda$) is covariant, i.e.,  ${\bm\Gamma}[\U_\alpha\bullet\U_\alpha^{\dagger}]=\W_\alpha{\bm\Gamma}[\bullet]\W_\alpha^{\dagger}$, with $\W_\alpha$ unitary. In this case, each term in \eqref{febf} is proportional to $\sum_{k,l}|\bra{\phi} \Lambda_k^T \otimes \T_\alpha\W_\alpha\Gamma_l\ket{\phi}|^2$, and can be independently optimized, as $\T_\alpha \W_\alpha$ is another unitary without constraints. Furthermore, the noise in the source can now be absorbed into the noise in the channel, $\bra{\phi} \Lambda_k^T \otimes \T_\alpha\W_\alpha\Gamma_l\ket{\phi} = \bra{\phi} \openone \otimes \T_\alpha\W_\alpha\Gamma_l \Lambda_k \ket{\phi}$, resembling the classical communication paradigm.

Further insight is also possible for weak interactions with the environments. Under this assumption, one expects that the initial  state is only slightly perturbed. Thus   $\bm\Lambda^{T}\otimes \bm\Gamma[\proj{\phi_{\U_{\alpha}}}]\approx (1-\epsilon) \proj{\phi_{\U_{\alpha}}} + \epsilon \rho_{\U_\alpha}$ is  a good approximation, with $\epsilon \ll 1$, and $\rho_{\U_\alpha}$ a state which depends on the initial state and channels. Equation~\eqref{febf} then becomes:
\begin{align}
\F_{\max}=\frac{1}{n^{2}}\!\max_{\{\U_\alpha,\T_\alpha\}}\sum_{\alpha} (1-\epsilon) |\bra{\phi} \openone \otimes\T_\alpha \U_\alpha \ket{\phi}|^2 \nonumber\\
 +  \epsilon \big<\phi_{\T_{\alpha}^{T}}\big| \rho_{\U_\alpha^*} \big|\phi_{\T_{\alpha}^{T}}\big>.
\label{convexChannel}
\end{align}
As $\epsilon\ll 1$, the best strategy is to maximize the first term in Eq.~\eqref{convexChannel}, leading to $\T_\alpha = \U_\alpha^\dagger$. This prescription is the same as for the STP (see caption of Fig.\ref{fig:diagrams}) -- as expected from the limiting case of no noise. One difference should however be pointed out: the choice of Alice's measurement basis (and hence of Bob's operations) is no longer inconsequential. The noise action might  break the equivalence among the bases,  defining a preferred direction. In Eq.~\eqref{convexChannel} this is easily seen by the possibility of maximizing the second term with an appropriate choice of $\{U_\alpha\}$.

In fact, the latter is also true for some relevant noise scenarios, for which the optimization in \eqref{febf} can be explicitly carried out. For example, it is easy to show that when ${\bm\Lambda}$ and/or ${\bm\Gamma}$ represent computational errors (bit-flip, phase-flip, or bit-phase-flip) or the interaction with a zero temperature reservoir~\cite{nielsen}, the optimal protocol will have $\T_\alpha = \U_\alpha^\dagger$, and the maximum fidelity can be obtained, for instance, setting $\{\U_{\alpha}\}=\{\openone,\sigma_{x},\sigma_{y},\sigma_{z}\}$, corresponding to a STP.  Not all the choices of $\{\U_\alpha\}$, however, lead to the best fidelity.


\emph{Numerics.}  It is clear from the discussion above that the protocol here introduced is qualitatively better than any other teleportation protocol to date. Now we set out to quantify the gain in taking into account the noise in the input state distribution. Below we numerically optimize Eq.\eqref{febf}, corresponding to the protocol restricted to unitaries and projective measurements (\emph{ii}), specialized to a system of qubits ($\dim \mathcal{H}_i=2$, for $i=A, B, \rm{in}$). To emphasize the importance of considering the effects of noise on the input state, we compare our realistic protocol with the one proposed by  Albeverio, Fei and Yang in Ref.~\cite{Albeverio:2002ys} (hereafter denoted by AFY protocol). The latter  was intended to noisy quantum channels and pure input states. Since the AFY does not require optimization over the measurement basis, we randomly choose different maximally entangled basis and apply the protocol to realistic situations where $\bm\Lambda\neq\openone$. The optimization is performed with the genetic algorithm routine GENMin~\cite{Tsoulos:2008fk}.

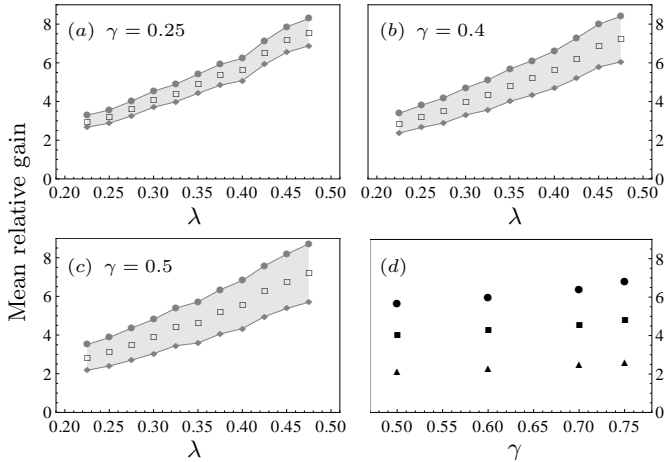
\begin{figure}[t!]
\begin{flushright}
\includegraphics[width= 0.98\columnwidth]{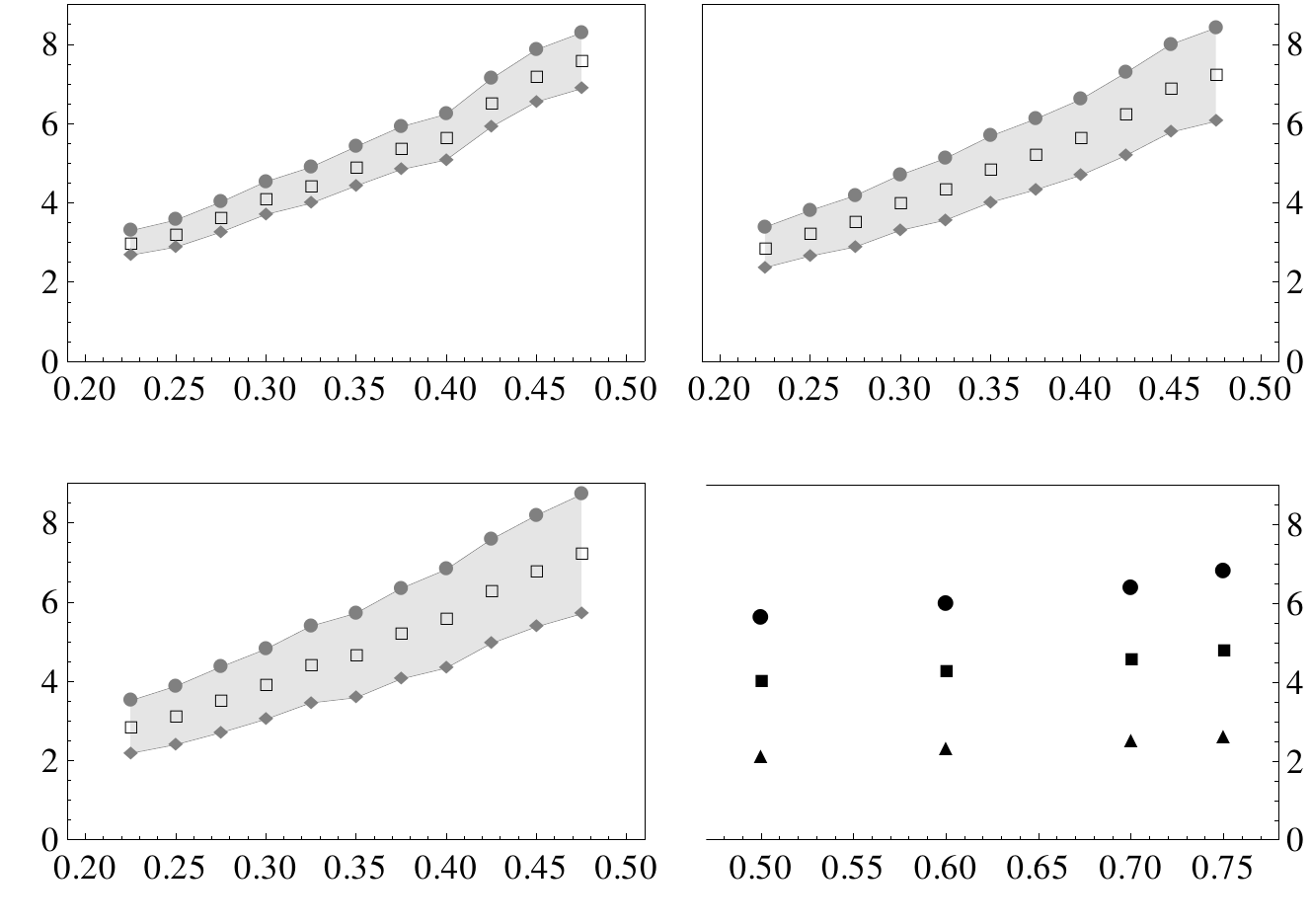}
\end{flushright}
\begin{picture}(0,0)
\put(-122,70){\rotatebox{90}{$\mbox{Mean relative gain}$}}
\put(65,20){$\gamma$}
\put(-55,19){$\lambda$}
\put(-55,107){$\lambda$}
\put(65,107){$\lambda$}
\put(-100,178){\begin{scriptsize}$(a)\,\,\,\gamma=0.25$ \end{scriptsize}}
\put(17,178){\begin{scriptsize}$(b)\,\,\, \gamma=0.4$ \end{scriptsize}}
\put(-100,89){\begin{scriptsize}$(c)\,\,\,\gamma=0.5$ \end{scriptsize}}
\put(17,89){\begin{scriptsize}$(d)$ \end{scriptsize}}
\end{picture}
\vspace{-0.9cm}
\caption{{\bf Random channels scenario.} Realistic protocol relative gain over AFY for randomly generated noisy scenarios. $(a)-(c)$ The shaded region shows the relative gain range for different measurement basis. Markers indicate the relative gain over AFY using the best measurement basis (diamonds), the worse (circles) and the average relative gain (squares) [relative gain defined as $100 (\overline{f}_{\max} - \overline f_{\text{AFY}})/\overline{f}_{\max}$]. $(d)$ Average relative gain in respect to $\text{AFY}_\text{max}$ for $\lambda=0.225$ (triangles), $0.375$ (squares), $0.475$ (circles) as a function of $\gamma$. \label{fig:bins}}
\end{figure}

We first address the scenario where both channel and input states are subjected to different, randomly generated, noisy processes~\cite{notaX}. By considering channels with a given strength, the typical relative gain of the realistic protocol can be determined by optimizing Eq.\eqref{febf} for many different random channel configurations. We gauge the strength of $\bm\Gamma$ by the amount of entanglement loss of the quantum resource when compared to the perfect channel: $\gamma(\bm\Gamma)=1-\mbox{Neg}(\Gamma[\proj{\phi}])$, with $\mbox{Neg}$ an entanglement measure~\cite{WernerVidal}. Likewise, for the noise  $\bm\Lambda$ on the input states, we use the fidelity loss $\lambda(\bm\Lambda)=1-\overline{\bra\psi\bm\Lambda\left[\proj\psi\right]\ket\psi}$, averaged over the set of pure input states. For the numerical investigations, we generated channels with strength parameters within intervals of length $0.01$. See Fig.\ref{fig:bins} for the results. 

It is clear from these results that, independent of the amount of entanglement in the resource state, the stronger $\bm\Lambda$ is the greater is the advantage of taking it into account. Furthermore, out of a sample of 38620 random noise configurations tested, in only $\sim$4.6\% of the instances our realistic protocol could be classically simulated ($\overline{f}_{\max}<2/3$). For the AFY protocol $\sim$25\% of the cases gave an average fidelity below the classical threshold of 2/3. As expected the weaker the noise  on the resource state is (smaller $\gamma$'s), the smaller is the difference between the AFY protocols, as the influence of carefully choosing the measurement basis is reduced. Additionally, having more quantum correlations at it's disposal, the realistic protocol can achieve bigger relative gains (shown in Fig.\ref{fig:bins}d for three values of $\lambda$).

Second, we compared the protocols when all the qubits are under the influence of identical bit-flip maps --- ${\bf \mathcal{E}}_{\rm{BP}}[\bullet] = (1-p) \bullet + p \,\sigma_x \bullet \sigma_x$, with $0\le p \le 1/2$. In this scenario, the realistic protocol gives fidelity above the classical threshold for all range of $p$. The relative gain of the realistic protocol against the average AFY increases monotonically with the noise strength, with a gain of 5.8\% at $p=0.25$, where the average AFY fidelity reaches the classical boundary. As mentioned previously, for this case the STP is already the best protocol. This was observed in  our numerical experiment with all three protocols, STP, AFY$_{\max}$ (with an optimal choice for Alice's basis), and our realistic protocol, yielding the same maximum average fidelity. In addition, we generated close to 10000 numerical experiments with ${\bm\Lambda}$ and ${\bm\Gamma}$ representing computational errors, finite-temperature reservoirs or compositions of these~\cite{nielsen}. These showed that as long as the STP outperforms any classical strategy, it reaches the optimal fidelity of Eq.\eqref{febf}, suggesting that, within the restricted set of operations considered, the STP is a robust protocol against the aforementioned decoherence processes.


\emph{Conclusions.} Teleportation spots yet another trait of quantum communications: quantum information is disturbed by the environment even before its transmission, and this disturbance cannot in general be accommodated as a faulty communication channel. Recognizing this is not only of  conceptual importance, but  has also practical implications. The  teleportation protocol here proposed appeals to this mindset shift in order to obtain sizable gains in communication quality.

\begin{acknowledgments}
We would like to acknowledge Fernando Brand\~ao, Rafael Chaves, Mark Fannes and Jeroen Wouters for fruitful discussions. B.G.T.  also thanks CAPES/DAAD PROBRAL program, the Quantum Optics and Statistics group at Freiburg University and the Institute for Theoretical Physics at K. U. Leuven. F.d-M. was supported by Alexander von Humboldt Foundation, and Belgian Interuniversity
Attraction Poles Programme P6/02. B.G.T. and R.M.F. were supported  by the Brazilian agencies CNPq, CAPES, FAPERJ and INCT-IQ.
\end{acknowledgments}




\begin{thebibliography}{10}
\bibitem{nota} We parametrize Alice and Bob operations via separable operations, for those include all the LOCC's. The physical process is, however, restricted to the LOCC subset.
\end{thebibliography}
\end{document}